\documentclass[12pt]{article}
\usepackage{graphicx}
\usepackage{lscape}
\usepackage{citesort}
\usepackage{amssymb}
\usepackage{appendix}
\usepackage{multirow}

\begin{document}
\def\qq{\langle \bar q q \rangle}
\def\uu{\langle \bar u u \rangle}
\def\dd{\langle \bar d d \rangle}
\def\sp{\langle \bar s s \rangle}
\def\GG{\langle g_s^2 G^2 \rangle}
\def\Tr{\mbox{Tr}}
\def\figt#1#2#3{
        \begin{figure}
        $\left. \right.$
        \vspace*{-2cm}
        \begin{center}
        \includegraphics[width=10cm]{#1}
        \end{center}
        \vspace*{-0.2cm}
        \caption{#3}
        \label{#2}
        \end{figure}
	}
	
\def\figb#1#2#3{
        \begin{figure}
        $\left. \right.$
        \vspace*{-1cm}
        \begin{center}
        \includegraphics[width=10cm]{#1}
        \end{center}
        \vspace*{-0.2cm}
        \caption{#3}
        \label{#2}
        \end{figure}
                }

\def\ds{\displaystyle}
\def\beq{\begin{equation}}
\def\eeq{\end{equation}}
\def\bea{\begin{eqnarray}}
\def\eea{\end{eqnarray}}
\def\beeq{\begin{eqnarray}}
\def\eeeq{\end{eqnarray}}
\def\ve{\vert}
\def\vel{\left|}
\def\ver{\right|}
\def\nnb{\nonumber}
\def\ga{\left(}
\def\dr{\right)}
\def\aga{\left\{}
\def\adr{\right\}}
\def\lla{\left<}
\def\rra{\right>}
\def\rar{\rightarrow}
\def\lrar{\leftrightarrow}  
\def\nnb{\nonumber}
\def\la{\langle}
\def\ra{\rangle}
\def\ba{\begin{array}}
\def\ea{\end{array}}
\def\tr{\mbox{Tr}}
\def\ssp{{\Sigma^{*+}}}
\def\sso{{\Sigma^{*0}}}
\def\ssm{{\Sigma^{*-}}}
\def\xis0{{\Xi^{*0}}}
\def\xism{{\Xi^{*-}}}
\def\qs{\la \bar s s \ra}
\def\qu{\la \bar u u \ra}
\def\qd{\la \bar d d \ra}
\def\qq{\la \bar q q \ra}
\def\gGgG{\la g^2 G^2 \ra}
\def\q{\gamma_5 \not\!q}
\def\x{\gamma_5 \not\!x}
\def\g5{\gamma_5}
\def\sb{S_Q^{cf}}
\def\sd{S_d^{be}}
\def\su{S_u^{ad}}
\def\sbp{{S}_Q^{'cf}}
\def\sdp{{S}_d^{'be}}
\def\sup{{S}_u^{'ad}}
\def\ssp{{S}_s^{'??}}

\def\sig{\sigma_{\mu \nu} \gamma_5 p^\mu q^\nu}
\def\fo{f_0(\frac{s_0}{M^2})}
\def\ffi{f_1(\frac{s_0}{M^2})}
\def\fii{f_2(\frac{s_0}{M^2})}
\def\O{{\cal O}}
\def\sl{{\Sigma^0 \Lambda}}
\def\es{\!\!\! &=& \!\!\!}
\def\ap{\!\!\! &\approx& \!\!\!}
\def\ar{&+& \!\!\!}
\def\ek{&-& \!\!\!}
\def\kek{\!\!\!&-& \!\!\!}
\def\cp{&\times& \!\!\!}
\def\se{\!\!\! &\simeq& \!\!\!}
\def\lth{\!\!\! &\le& \!\!\!}
\def\gth{\!\!\! &\ge& \!\!\!}
\def\eqv{&\equiv& \!\!\!}
\def\kpm{&\pm& \!\!\!}
\def\kmp{&\mp& \!\!\!}
\def\mcdot{\!\cdot\!}
\def\erar{&\rightarrow&}


\def\simlt{\stackrel{<}{{}_\sim}}
\def\simgt{\stackrel{>}{{}_\sim}}


\title{
         {\Large
                 {\bf
More about on the mass of the new charmonium states
                 }
         }
      }

\author{\vspace{1cm}\\
{\small T. M. Aliev \thanks {
taliev@metu.edu.tr}~\footnote{Permanent address: Institute of
Physics, Baku, Azerbaijan.}\,\,, H. \"{O}z\c{s}ahin \thanks {e-mail:
hikmet.ozsahin@metu.edu.tr}\,\,, 
M. Savc{\i} \thanks
{savci@metu.edu.tr}} \\
{\small Physics Department, Middle East Technical University,
06531 Ankara, Turkey }}

\date{}

\begin{titlepage}
\maketitle
\thispagestyle{empty}

\begin{abstract}

If the $X(3872)$ is described by the picture as a mixture of the charmonium
and molecular $D^{\ast} D$ states; $Y(3940)$ as a mixture of the
$\chi_{c0}$ and $D^\ast D^\ast$ states; and $X(4260)$ as a mixture of the
tetra-quark and charmonium sates, their orthogonal combinations should also
exist. We estimate the mass and residues of the states within the QCD sum
rules method. We find that the mass splitting among $X$, $Y$ and their
orthogonal states is at most $200~MeV$. Experimental search of these new
states can play critical role for establishing the nature of the new
charmonium states.

\end{abstract}

~~~PACS numbers: 11.55.Hx, 12.39.Mk, 12.39.--x, 14.40.Rt
\end{titlepage}

\section{Introduction}

Last decade was quite an exciting and productive period in particle physics.
Starting from the first observation of the $X(3872)$ meson by the Belle
Collaboration \cite{Rboz01} up to now more than twenty new charmonium states have been
observed (for a review see \cite{Rboz02,Rboz03}).The main lesson from these
discoveries is that these states (referred as $XYZ$ states) can not be
described by the simple quark model, i.e., as a quark-antiquark system, and
they are believed to have more complex structures. Understanding the
structure and dynamics of these states is one of the hot problems in
particle physics. In investigating the properties of these states two
pictures are widely used, namely, four-quark (tetra quark) or meson
molecules (bound states of two mesons) (for a review see \cite{Rboz03} and
\cite{Rboz04}).

The mass and some of the strong coupling constants of the XYZ mesons with
light mesons are widely studied within the QCD sum rules method
\cite{Rboz05} in many works (see for example \cite{Rboz06,Rboz07}).  
The first calculation of the mass of the $X(3872)$ meson as a tetra-quark
state with the quantum numbers $J^{PC}=1^{++}$ is performed within QCD sum
rules method in \cite{Rboz06}. The $X(3872)$ meson as a $D^\ast D$ molecular
state with the interpolating current,
\bea
\label{eboz01}
j_\mu^{mol} = {1\over \sqrt{2}} \Big[(\bar{q}_a \gamma_5 c_a) ( \bar{c}
\gamma_\mu q_b) - (\bar{q}_a \gamma_\mu c_a) ( \bar{c} \gamma_5 q_b)\Big]~,
\eea
was investigated in \cite{Rboz07}. The next step in analysis of the
properties of $X(3872)$ is that it is assumed to be a mixture of the
charmonium and $D^{\ast} D$ molecules \cite{Rboz08}. The interpolating
current has the form,
\bea
\label{eboz02}
j_\mu^{(1)} = \cos\theta_1 j_\mu^{ch} + \sin\theta_1 j_\mu^{mol}~,
\eea
where
\bea
\label{nolabel01}
j_\mu^{ch} = {1\over 6 \sqrt{2}} \qq (\bar{c} \gamma_\mu\gamma_5 c)~, \nnb
\eea
and $j_\mu^{mol}$ is given in Eq. (\ref{eboz01}).
 In \cite{Rboz08},
in order to reproduce the
experimental value of the mass, the mixing angle is
calculated to be $\theta_1=(9 \pm 4)^0$.

As far as the the state $Y(3940)$ is concerned, it is described by
the mixture of the scalar $\chi_{c0}$ and $D^\ast D^\ast$ molecules
\cite{Rboz09}. In
other words its interpolating current can be written as,
\bea
\label{eboz03}
j^{(2)} = \cos\theta_2 \Bigg({- \qq \over \sqrt{2}} \bar{c}c \Bigg)
+\sin\theta_2(\bar{q}
\gamma_\mu c)(\bar{c}\gamma^\mu q)~.
\eea
In the same manner, the mixing angle is calculated by requiring that it
should reproduce the mass of the $Y(3940)$ state, and is found to have the
value $\theta_2=(76\pm 5)^0$.

Finally, if one assumes that $Y(4260)$ state is a
mixture of the tetra quark and the charmonium current given as \cite{Rboz10},
\bea
\label{eboz04}
j_\mu^{(3)} = \cos\theta_3 j_\mu^{ch} + \sin\theta_3 j_\mu^{tet}~,
\eea
where
\bea
\label{eboz05}
j_\mu^{ch} \es {1\over \sqrt{2}} \qq (\bar{c} \gamma_\mu c)~, \nnb \\
j_\mu^{tet} \es  {1\over \sqrt{2}} \varepsilon^{abc} \varepsilon^{dec} \Big[
(q_a^T C \gamma_5 c_b) (\bar{q}_d \gamma_\mu \gamma_5 C \bar{c}_e^T)
+ (q_a^T C \gamma_\mu \gamma_5 c_b) (\bar{q}_d \gamma_5 C
\bar{c}_e^T) \Big]~,
\eea 
and  requires that it should reproduce the mass $Y(4260)$
state, the mixing angle is calculated to be $\theta_3 = (53 \pm 5)^0$
\cite{Rboz10}.

Using the same pictures the corresponding mixing angles are estimated within
the QCD sum rules method in \cite{Rboz11}, whose values are calculated to
have the values $\theta_1=(2.4\pm 0.6)^0$, $\theta_2=(20\pm 2)^0$, and
$\theta_3=(20\pm 3)^0$, respectively.

If the correct pictures of the $X(3872)$, $Y(3940)$ and $Y(4260)$
states were mixture of the molecular or tetra quark states with charmonium
states, the orthogonal combinations of these states should also exist. In
other words, these orthogonal $X^\prime$, $Y^\prime$ and $Z^\prime$ states
should be described by the following interpolating currents:
\bea
\label{eboz06}
j_\mu^{(1)} \es - \sin\theta_1 j_\mu^{ch} + \cos\theta_1 j_\mu^{mol}~, \\
\label{eboz07}
j^{(2)} \es - \sin\theta_2 \Bigg({\qq \over \sqrt{2}} \bar{c}c\Bigg)
+\cos\theta_2(\bar{q}       
\gamma_\mu c)(\bar{c}\gamma^\mu q)~. \\
\label{eboz08}
j_\mu^{(3)} \es - \sin\theta_3 j_\mu^{ch} + \cos\theta_3 j_\mu^{mol}~,      
\eea

The aim of this letter is to calculate the mass and residues of these states
described by the interpolating currents (\ref{eboz06}),
(\ref{eboz07}) and (\ref{eboz06}).

Consider the following two-point correlation functions,
\bea
\label{eboz09}
\Pi_{\mu\nu}^{(1)} \es i \int d^4x e^{iqx} \lla 0 \vel T\Big\{ j_\mu^{(1)}
(x) j_\nu ^{(1)} (0) \Big\} \ver 0 \rra~\nnb \\
\es \Pi_1^{(1)} (q^2) \Bigg( g_{\mu\nu} - {q_\mu q_\nu \over q^2} \Bigg) +
\Pi_0^{(1)} {q_\mu q_\nu \over q^2}~,\\
\label{eboz10}
\Pi^{(2)} \es i \int d^4x e^{iqx}  \lla 0 \vel T\Big\{ j^{(2)} (x) j^{(2)}
(0) \Big\} \ver 0 \rra~\nnb \\
\es \Pi^{(2)} (q^2)~, \\
\label{eboz11}
\Pi_{\mu\nu}^{(3)} \es i \int d^4x e^{iqx} \lla 0 \vel T\Big\{
j_\mu^{(3)}(x) j_\nu ^{(3)} (0) \Big\} \ver 0 \rra~\nnb \\
\es \Pi_1^{(3)} (q^2) \Bigg( g_{\mu\nu} - {q_\mu q_\nu \over q^2}\Bigg) +
\Pi_0^{(3)} {q_\mu q_\nu \over q^2}~.
\eea
According to the duality principle the phenomenological part
of these  correlation functions can be
calculated in terms of quarks, gluons and hadrons by inserting complete set
of hadrons carrying the same quantum quantum numbers as the interpolating
currents themselves. Then isolating the ground states, and performing
summation over the spins (if ever exist) we get,
\bea
\label{eboz12}
\Pi_{\mu\nu}^{(1)} \es {\la 0 \ve j_\mu^{(1)} \ve B(q) \ra \la B(q) \ve
j_\nu^{(1)} \ve 0 \ra \over q^2-m_{B_1}^2 } + \cdots \nnb \\
\es {\lambda^{(1)2} \over m_{B_1}^2 - q^2} \Bigg( - g_{\mu\nu} + {q_\mu q_\nu
\over m_{B_1}^2 } \Bigg) + \cdots\\
\label{eboz13}
\Pi^{(2)} \es {\lambda^{(2)2} \over m_{B_2}^2 - q^2}+ \cdots \\ 
\label{eboz14}
\Pi_{\mu\nu}^{(3)} \es 
{\lambda^{(3)2} \over m_{B_3}^2 - q^2} \Bigg( - g_{\mu\nu} + {q_\mu
q_\nu \over m_{B_1}^2 } \Bigg) + \cdots~,
\eea 
where $\cdots$ means the contributions of the higher states and continuum,
and we have used,
\bea
\label{nolabel02}
\lla 0 \vel j_\mu^{(1),(3)} \ver B(q) \rra \es
\lambda^{(1),(3)} \varepsilon_\mu^{(1),(3)}~, \nnb \\
\lla 0 \vel j^{(2)} \ver B(q) \rra \es
\lambda^{(2)}~. \nnb
\eea
The continuum contribution to the spectral density is modeled as a spectral
density from the operator product expansion (OPE) starting form some
threshold $s_0$, i.e.,
\bea
\label{nolabel03}
\rho^{cont} (s) \es \rho^{OPE} \theta(s-s_0)~,\nnb
\eea
where $\theta(s-s_0)$ is the Heaviside step function.

In order to construct the sum rules for the mass and residues of new
charmonium states the calculation of
the correlation functions in terms of the quark and gluon
degrees of freedom using the operator product expansion (OPE) are needed.
On the other hand, to be able to
calculate the correlation functions from QCD side the
heavy and light quark propagators are needed, whose
expressions in the coordinate space are given as,
\bea
\label{nolabel04}
S_q(x) \es {i \rlap/{x} \over 2 \pi^2 x^4} - {m_q \over 4 \pi^2 x^2} - {\qq 
\over 12} \Bigg( 1- {im_q \over 4} \rlap/{x} \Bigg) - {x^2 \over 192} m_0^2
\qq \Bigg( 1- {im_q \over 6} \rlap/{x} \Bigg) \nnb \\
\ek {\alpha_s \pi x^2 \rlap/{x} \over 2^3 3^5} \qq^2 + {1\over 32 \pi^2} g_s
G_{\mu\nu} {\sigma^{\mu\nu} \rlap/{x} + \rlap/{x} \sigma^{\mu\nu} \over x^2}
- {\pi^2 x^4 \over 2^8 3^3} \qq \GG \nnb \\
\ek {\alpha_s m_q \pi x^4 \over 2^5 3^5} \qq^2
- {m_q x^2 \rlap/{x} \over 2^7 3^2} m_0^2 \qq + \cdots \nnb \\ \nnb \\
S_Q \es {m_Q^2 \over 4 \pi^2} \Bigg[
{i \rlap/{x} \over (-x^2)} K_2 (m_Q \sqrt{-x^2}) + {1\over (\sqrt{-x^2})} 
K_1 (m_Q \sqrt{-x^2}) \Bigg] \nnb \\
\ek { m_Q g_s G_{\mu\nu} \over 32 \pi^2 }
\Bigg[ {i(\sigma^{\mu\nu} \rlap/{x} + \rlap/{x} \sigma^{\mu\nu}) \over
(\sqrt{-x^2})} K_1 (m_Q \sqrt{-x^2}) + 2  \sigma^{\mu\nu} K_0 (m_Q
\sqrt{-x^2})\Bigg] \nnb \\
\ek {\GG \over 2304 \pi^2} \Bigg[ (i m_Q \rlap/{x} - 6) (\sqrt{-x^2})
K_1 (m_Q \sqrt{-x^2}) + m_Q (-x^2) K_2 (m_Q \sqrt{-x^2}) \Bigg]~. \nnb
\eea

The theoretical part of the correlation function(s) can be written in terms
of the dispersion relation as,
\bea
\label{eboz15}
\Pi^{(i)} (q^2) = \int_{4 m_c^2}^\infty ds {\rho^{OPE^{(i)}} (s) \over s-q^2}~.
\eea
Choosing the coefficient of the $g_{\mu\nu}$ structure for the correlators
(\ref{eboz09}) and (\ref{eboz11}), for the spectral densities we get,
\bea
\label{eboz16}
\rho_1(q^2) \es
{1\over 4096 \pi^6} \int_{\alpha_{min}}^{\alpha_{max}}  d\alpha
\int_{\beta_{min}}^{\beta_{max}}  d\beta
\Big\{ 3 \alpha\beta [1-(\alpha+\beta)^2] \mu_1^4 \nnb \\
\ek 6 m_c m_q (\alpha+\beta) (1-\alpha-\beta) (3-\alpha+\beta) \mu_1^3 \nnb \\ 
\ek 48 \pi^2 \Big[ (\alpha+\beta) (1+\alpha+\beta) m_c
+ 2 \alpha \beta m_q \Big] \qq \mu_1^2 \nnb \\
\ar 24 \pi^2 \Big[ 16 m_c m_q + (\alpha+\beta) m_0^2  \Big] m_c \qq \mu_1
\Big\} \cos^2\theta  \nnb \\
\ar {1\over 1536 \pi^4} \Big\{
36 \alpha (1-\alpha) m_q \qq \mu_2^2 \cos^2\theta
- 2 \Big( m_0^2
\Big[ 9 m_c - 6 \alpha (1-\alpha) m_q \Big] \nnb \\
\ar  8 \pi^2 \alpha (1-\alpha) \qq \Big) \qq \mu_2 \cos^2\theta
+ 4 \Big(3 m_0^2 m_q \Big[ 3 m_c^2 - \alpha (1-\alpha) s \Big]
\nnb \\
\ar 2 \pi^2 \Big[ 10 m_c^2 - 9 m_c m_q + 2 \alpha (1-\alpha) s
\Big] \qq \Big)
\qq \cos^2\theta \nnb \\
\ek 16 \pi^2 \Big[ m_c^2 + \alpha (1-\alpha) (\mu_2-s)
\Big] \qq^2 \sin\theta ( \sin\theta + 2 \cos\theta) \Big\}~,
\\ \nnb \\
\label{eboz17}
\rho_2(q^2) \es
{1\over 512 \pi^6} \int_{\alpha_{min}}^{\alpha_{max}}  d\alpha
\int_{\beta_{min}}^{\beta_{max}}  d\beta 
\Big\{ - 3 \alpha\beta(1-\alpha-\beta) \mu_1^4 +
6 (1-\alpha-\beta) (\alpha+\beta) m_c m_q \mu_1^3 \nnb \\
\ar 24 \pi^2 (\alpha+\beta) m_c \qq \mu_1^2  
- 192  \pi^2 m_c^2 m_q \qq \mu_1 \Big\}\cos^2\theta  \nnb \\
\ar {1\over 384 \pi^4} \int_{\alpha_{min}}^{\alpha_{max}}  d\alpha
\Big\{ -36 \alpha (1-\alpha) m_q \qq \mu_2^2 \ \cos^2\theta \nnb \\
\ar \Big( 3 m_0^2 \Big[ 3 m_c - 8 \alpha (1-\alpha) m_q\Big] + 64 \pi^2
\alpha (1-\alpha) \qq \Big) \qq\mu_2 \cos^2\theta \nnb \\
\ar \Big( 6 m_0^2 m_q \Big[ -6 m_c^2 + 2\alpha (1-\alpha) s\Big] + 16 \pi^2
\Big[ m_c (3m_q - 4 m_c) - 2 \alpha (1-\alpha) s \Big] \qq \Big) \qq\cos^2\theta \nnb \\
\ar 48 \pi^2 \Big[m_c^2 + \alpha (1-\alpha) (2 \mu_2-s) \Big] \qq^2
\sin\theta(3 \sin\theta - 2 \sqrt{2} \cos\theta) \Big\}~, \\ \nnb \\
\label{eboz18}
\label{eboz18}
\rho_3(q^2) \es
{1\over 3072 \pi^6} \int_{\alpha_{min}}^{\alpha_{max}}  d\alpha
\int_{\beta_{min}}^{\beta_{max}}  d\beta
\Big\{ 3 \alpha\beta [1-(\alpha+\beta)^2] \mu_1^4 - 2 (1-\alpha-\beta)^3
m_c^2 \mu_1^3 \nnb \\ 
\ar 192 \pi^2 \alpha\beta m_q \qq \mu_1^2
- 96 \pi^2 (5-\alpha-\beta) m_c^2 m_q \qq \mu_1 
- 16 \pi^2 m_0^2
m_c^2 m_q \qq \Big\} \cos^2\theta  \nnb \\
\ar {1\over 192 \pi^4} \int_{\alpha_{min}}^{\alpha_{max}}  d\alpha
\Big\{ \alpha (1-\alpha) \qq \mu_2 \Big[6 m_q \mu_2 - (m_0^2 m_q + 16 \pi^2
\qq \Big] \cos^2\theta \nnb \\
\ek \Big( 2 m_0^2 m_q \Big[ 3 m_c^2 + \alpha (1-\alpha) s\Big] + 8 \pi^2 \Big[
m_c^2 + \alpha (1-\alpha) (m_0^2-s) \Big] \qq \Big) \qq \cos^2\theta \nnb \\
\ar 24 \pi^2 \Big[m_c^2 - \alpha (1-\alpha) (\mu_2-s) \Big] \qq^2 
\sin\theta(3 \sin\theta - \cos\theta) \Big\}~,
\eea
where
\bea
\label{nolabel05}
\mu_1 \es {m_c^2 (\alpha+\beta) \over \alpha\beta} -s~,\nnb \\
\mu_2 \es \mu_1 (\beta \to 1- \alpha)~,\nnb \\
\beta_{min} \es {\alpha m_c^2\over \alpha s - m_c^2}~, \nnb \\
\beta_{max} \es 1-\alpha~, \nnb \\
\alpha_{min} \es  {1\over 2} (1-v)~, \nnb \\
\alpha_{max} \es  {1\over 2} (1+v)~,~ \mbox{\rm and},\nnb \\
v \es \sqrt{1- {4 m_c^2 \over s}}~. \nnb 
\eea

For simplicity, in these expressions we do not present the terms
proportional with the gluon condensate $\GG$, and dimension-6 quark
condensate operators. It should be noted that these spectral densities are
all calculated  in \cite{Rboz08,Rboz09,Rboz10}. Perturbative parts of our
results coincide with those presented in \cite{Rboz08,Rboz09,Rboz10}, while
there are some discrepancies in quark condensates and $d=5$ operators.

Equating the spectral densities given in Eqs.(\ref{eboz16}), (\ref{eboz17}) and
(\ref{eboz18}) with the coefficients of
$\Pi_1$, $\Pi_2$ and $\Pi_3$, respectively,
and performing Borel transformation over $-q^2$, we get
\bea
\label{eboz19}
\lambda^{(i)2} e^{-m_1^2/M^2} \es \int_{4 m_c^2}^{s_0} ds \rho_i^{(1)}
e^{-s/M^2}~,
\eea
where $i=1,2,3$.
In deriving Eq.(\ref{eboz19}) the quark-hadron duality ansatz have been
used, i.e., the contribution of the continuum and higher states are assumed
to be same as the perturbative ones starting on from the threshold.

In order to obtain the mass sum rules for the mass, We take the derivative of
Eq. (\ref{eboz19}) with respect to $1/M^2$ on both sides, and divide the
obtained result by itself, from which we obtain
\bea
\label{eboz20}
  m_i^2 = {\ds \int_{4 m_c^2}^{s_0} ds \, s \rho_i^{(1)} e^{-s/M^2} \over
\ds \int_{4 m_c^2}^{s_0} ds \rho_i^{(1)} e^{-s/M^2} }~.
\eea

\section*{Numerical analysis}

Here in this section we present the results of the numerical analysis on the
mass and residues of of the considered new charmonium states. For the
$c$-quark masses we have used its $\bar{MS}$ scheme masses,
$\bar{m}_c(\bar{m}_c)= (1.28 \pm 0.03)~GeV$, $\qq
=(1~GeV)=-(0.246_{-19}^{+28}~MeV)^3$, $m_0^2 = (0.8 \pm 0,2)~GeV^2$,
$\sp = 0.8 \qq$, and $m_s(2~GeV)=(102 \pm 8)~MeV$.

The sum rules for the mass and residue contain two auxiliary variables,
namely, the Borel mass parameter $M^2$, and the continuum threshold $s_0$.
Any physical quantity should be independent of them, and therefore our
primary aim is to find the regions of $M^2$ and $s_0$, where the mass and
residue are practically independent of them. It should be noted here that
the continuum threshold itself is not a totally arbitrary
parameter, but related to the first excited state. However, since
no up to date information is available about the first excited states of
the considered charmonium states, we take the value of the continuum
threshold as $s_0=(m_{ground} + 0.5)^2~GeV^2$. The lower limit of $M^2$ is
determined by requiring that the operator product expansion series is
convergent. In other words the contribution of the perturbative part
must dominate over the nonperturbative one. The upper bound of $M^2$ is decide
from the condition that the contribution coming from the continuum constitutes
about $1/3$ of the contribution coming from the perturbative part, that is, the
ratio
\bea
\label{nolabel06}
R = {\int_{s_0}^\infty ds \rho(s) e^{-s/M^2} \over 
\int_{4 m_c^2}^\infty ds \rho(s) e^{-s/M^2}} \, \textless \, {1\over 3}~,
\nnb
\eea
which leads to the following ``working regions" of the Borel mass parameter,
\bea
\label{nolabel07}
2.4 \lth M^2 \le 4.0~GeV^2~\mbox{\rm for $X^\prime$ state}~, \nnb \\
2.5 \lth M^2 \le 4.0~GeV^2~\mbox{\rm for $Y^\prime$ state}~, \nnb \\
2.5 \lth M^2 \le 5.0~GeV^2~\mbox{\rm for $Y^{\prime\prime}$ state}~, \nnb
\eea
In figures (1), we present the dependence of the mass of $X^\prime$ state on
$M^2$, at the fixed value of the continuum threshold $\sqrt{s_0}=4.4~GeV$
and of the
mixing angle $\theta$. We observe from these figures that, in the considered
domains of $M^2$, the results exhibit good stability with respect to the
variation in
$M^2$, and seem to be practically insensitive to the variations in $s_0$ and
the mixing angle $\theta$. 
As has already been noted, if $X$ meson is represented as mixture of
the charmonium and $D^\ast D$ molecular states, it is found in
\cite{Rboz11} the mixing angle is equal to $\theta=(2.4\pm 0.3)^0$.
In other words, in this picture $X(3872)$ state can be said to be composed of pure
charmonium and its orthogonal combination $X^\prime$ which is described by
the $D^\ast D$ molecular state. The study of the decay channels of
$X(3872)$ and $X^\prime$ states can give unambiguous useful information
about the ``correct" pictures of these states. 
Our final result for the mass of the $X^\prime$ is that,
\bea
\label{nolabel08}
m_{X^\prime}= (3.75 \pm 0.15)~GeV~. \nnb
\eea

In Fig. (2) we present the $M^2$ dependence of the residue of $X^\prime$ on
$M^2$ for
the mixing angle $\theta = 2.4^0$, at the fixed value of the continuum
threshold $\sqrt{s_0}=4.4~GeV$. We observe that the residue is weakly
dependent on $M^2$, and we deduce from this figure that 
\bea
\label{nolabel081}
\lambda_{X^\prime} = (1.3 \times \pm 0.2)\times 10^{-2}~GeV^3.\nnb
\eea
 The dependencies of the mass and residue $Y\prime$ state on $M^2$ for the
mixing angle $\theta=20^0$, at $\sqrt{s_0}=4.4~GeV$ are presented in Figs.
(3) and (4), respectively. We observe from these figures that, again, the
mass and residue of $Y\prime$ state seem to be practically insensitive with
respect to the variation in $M^2$, whose values are calculated to be,
\bea
\label{nolabel09}
m_{Y^\prime} \es (3.85 \pm 0.20)~GeV~,\nnb \\
\lambda_{Y^\prime} \es (1.9 \pm 0.4)\times 10^{-2}~GeV^{-3}~.\nnb
\eea

Performing similar approach for the $Y^{\prime\prime}$ state, we see from
Figs. (5) and (6) that,
\bea
\label{nolabel09}
m_{Y^{\prime\prime}} \es (4.4 \pm 0.1)~GeV~,\nnb \\
\lambda_{Y^{\prime\prime}} \es (2.0 \pm 0.2)\times 10^{-2}~GeV^{-3}~.\nnb 
\eea

We would like to note here that, only $Y^{\prime\prime}$ state is sensitive
to the change in the value of the mixing angle $\theta$.
With the increasing value of $\theta$, the mass of the
$Y^{\prime\prime}$ state is decreasing, for example, at $\theta=40^0$
$m_{Y^{\prime\prime}} = (4.0\pm 0.1)~GeV$; and at $\theta=50^0$
$m_{Y^{\prime\prime}} = (3.9\pm 0.1)~GeV$.

From these predictions of the masses of $X^\prime$, $Y^\prime$ and
$Y^{\prime\prime}$ states, and the experimental values of $X(3872)$,
$Y(3940)$, $Y(4260)$ states we observe that the splittings among these
states are at most $100 \le \Delta m \le 200~MeV$. These results can be
checked in future planned experiments after the discovery of the $X^\prime$,
$Y^\prime$ and $Y^{\prime\prime}$ states.

The predicted results for the residues of the $Y^{\prime\prime}$ and 
$Y(4260)$; and $Y(3940)$ states  respectively are very close to each other,
while the result for the $X^\prime$ state is three 
times larger compared
to that of the $X(3872)$ state.

Using the picture that assumes $X(3872)$ as the mixture of the charmonium and
the molecular $D^{\ast} D$ states; $Y(3940)$ as the mixture of the
$\chi_{c0}$ and $D^\ast D^\ast$ states; and $X(4260)$ as the mixture of the
tetra-quark and charmonium states, we calculate the mass and the residues of
their orthogonal states within the QCD sum rules method. We obtain that the
mass splittings among these states is around $200~MeV$. Experimental search
of new orthogonal states can be quite useful in establishing correct picture
of these states. The study of the decay channels can also be very useful in
this investigation.

\newpage

\section*{Figure captions}
{\bf Fig. (1)} The dependence of the mass of the $X^\prime$ state on Borel
mass square $M^2$ for the mixing angle $\theta=2.4^0$, at the fixed value
of the continuum threshold $\sqrt{s_0}=4.4~GeV$. \\ \\
{\bf Fig. (2)} The dependence of the residue of the $X^\prime$ state on Borel 
mass square $M^2$ for the mixing angle $\theta=2.4^0$, at the fixed value   
of the continuum threshold $\sqrt{s_0}=4.4~GeV$. \\ \\
{\bf Fig. (3)} The dependence of the mass of the $Y^\prime$ state on Borel 
mass square $M^2$ for the mixing angle $\theta=20^0$, at the fixed value   
of the continuum threshold $\sqrt{s_0}=4.4~GeV$. \\ \\
{\bf Fig. (4)} The dependence of the residue of the $Y^\prime$ state on Borel 
mass square $M^2$ for the mixing angle $\theta=20^0$, at the fixed value   
of the continuum threshold $\sqrt{s_0}=4.4~GeV$. \\ \\
{\bf Fig. (5)} The dependence of the mass of the $Y^{\prime\prime}$ state on Borel 
mass square $M^2$ for the mixing angle $\theta=20^0$, at the fixed value   
of the continuum threshold $\sqrt{s_0}=4.6~GeV$. \\ \\
{\bf Fig. (6)} The dependence of the residue of the $Y^{\prime\prime}$ state on Borel 
mass square $M^2$ for the mixing angle $\theta=20^0$, at the fixed value   
of the continuum threshold $\sqrt{s_0}=4.6~GeV$.

\newpage

\begin{figure}
\vskip 3. cm
    \includegraphics{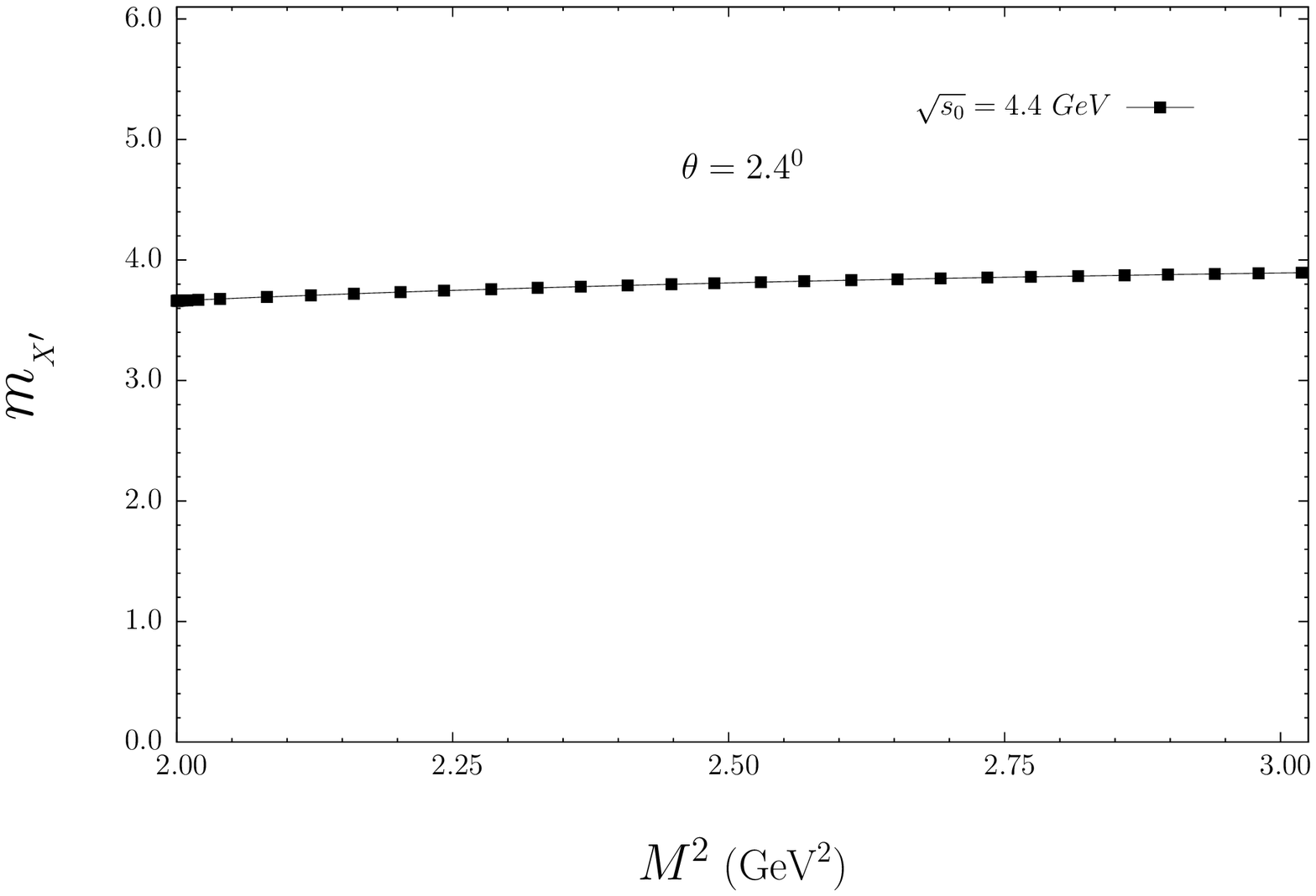}
\vskip 7.0cm
\caption{}
\end{figure}

\begin{figure}
\vskip 3. cm
    \includegraphics{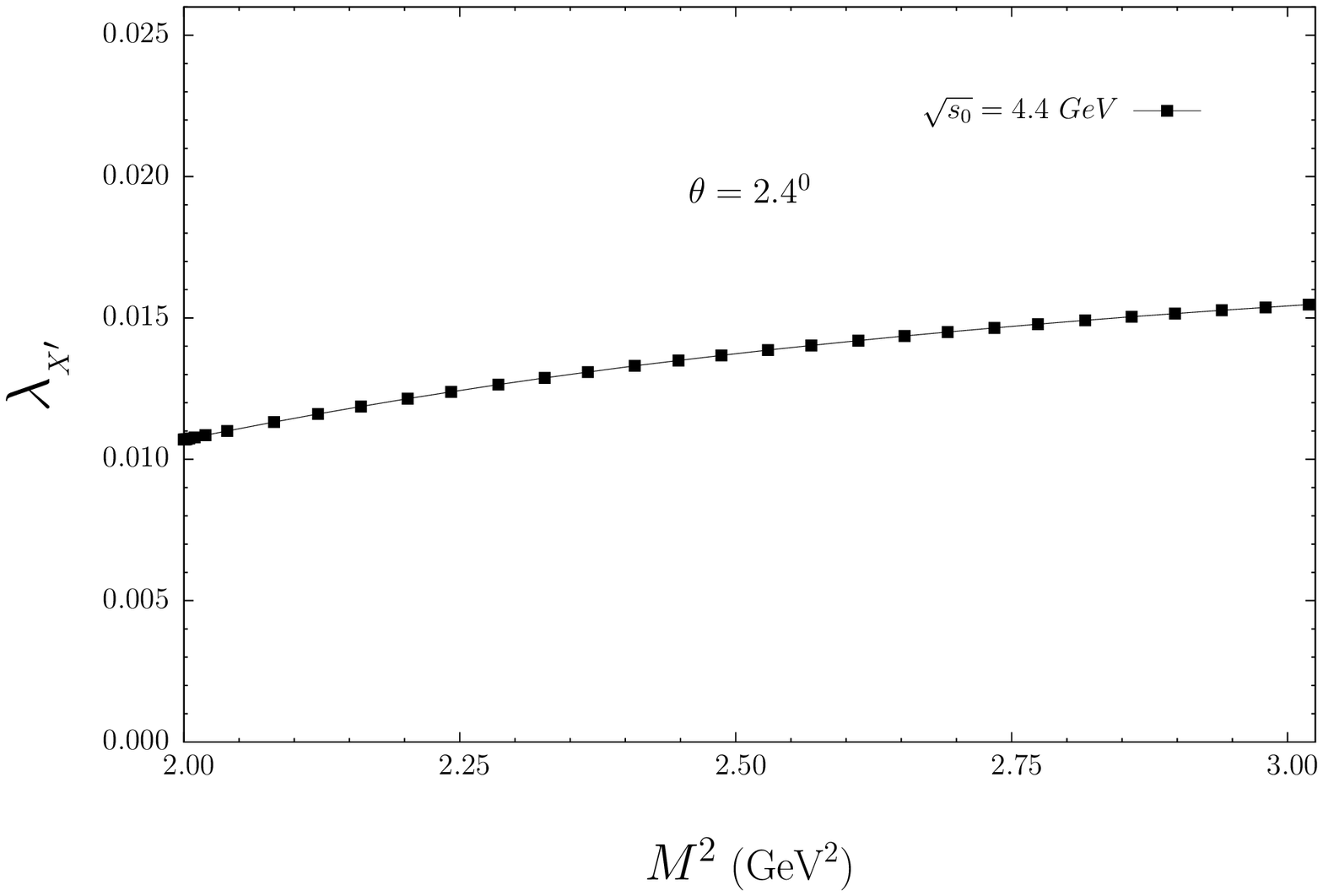}
\vskip 7.0cm
\caption{}
\end{figure}

\begin{figure}
\vskip 3. cm
    \includegraphics{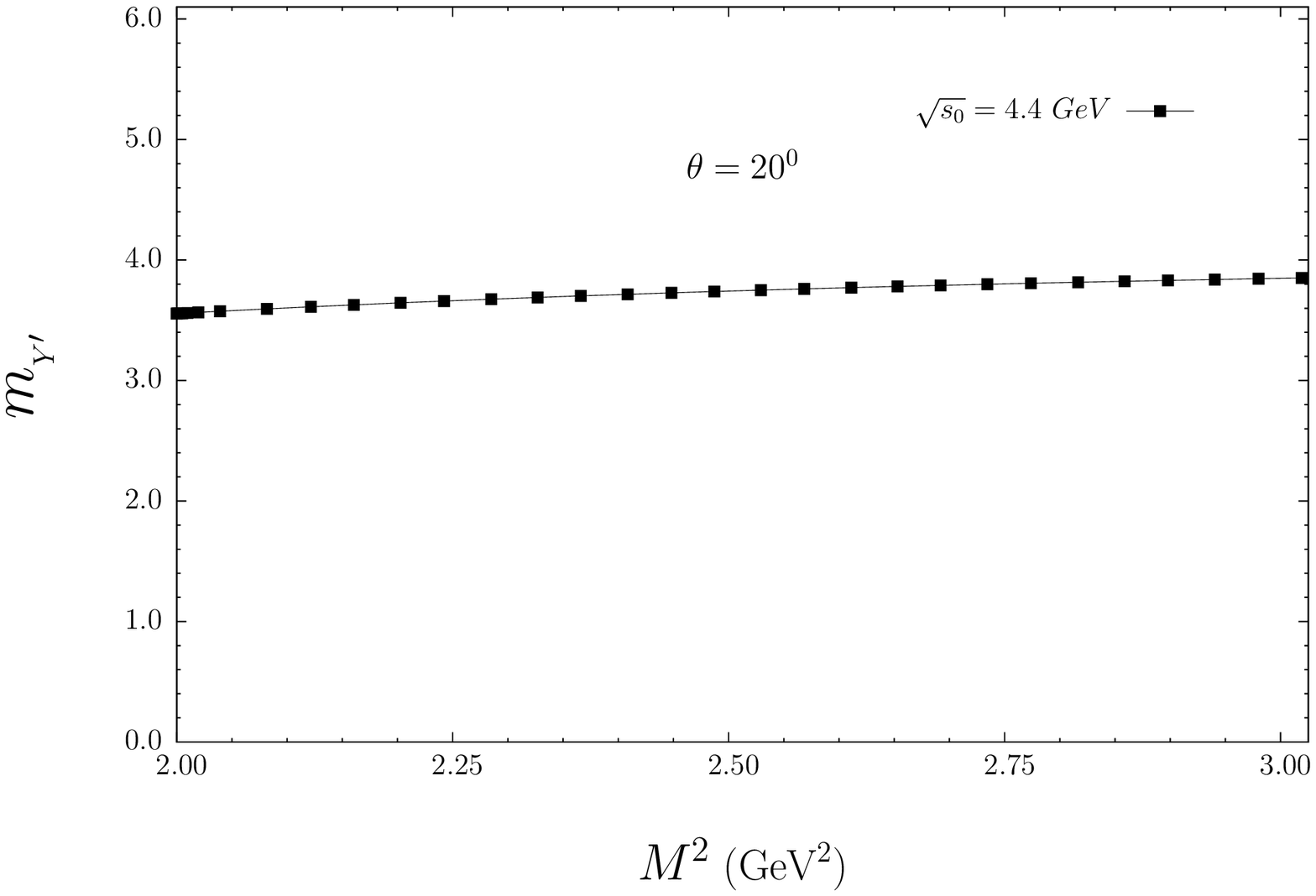}
\vskip 7.0cm
\caption{}
\end{figure}

\begin{figure}
\vskip 3. cm
    \includegraphics{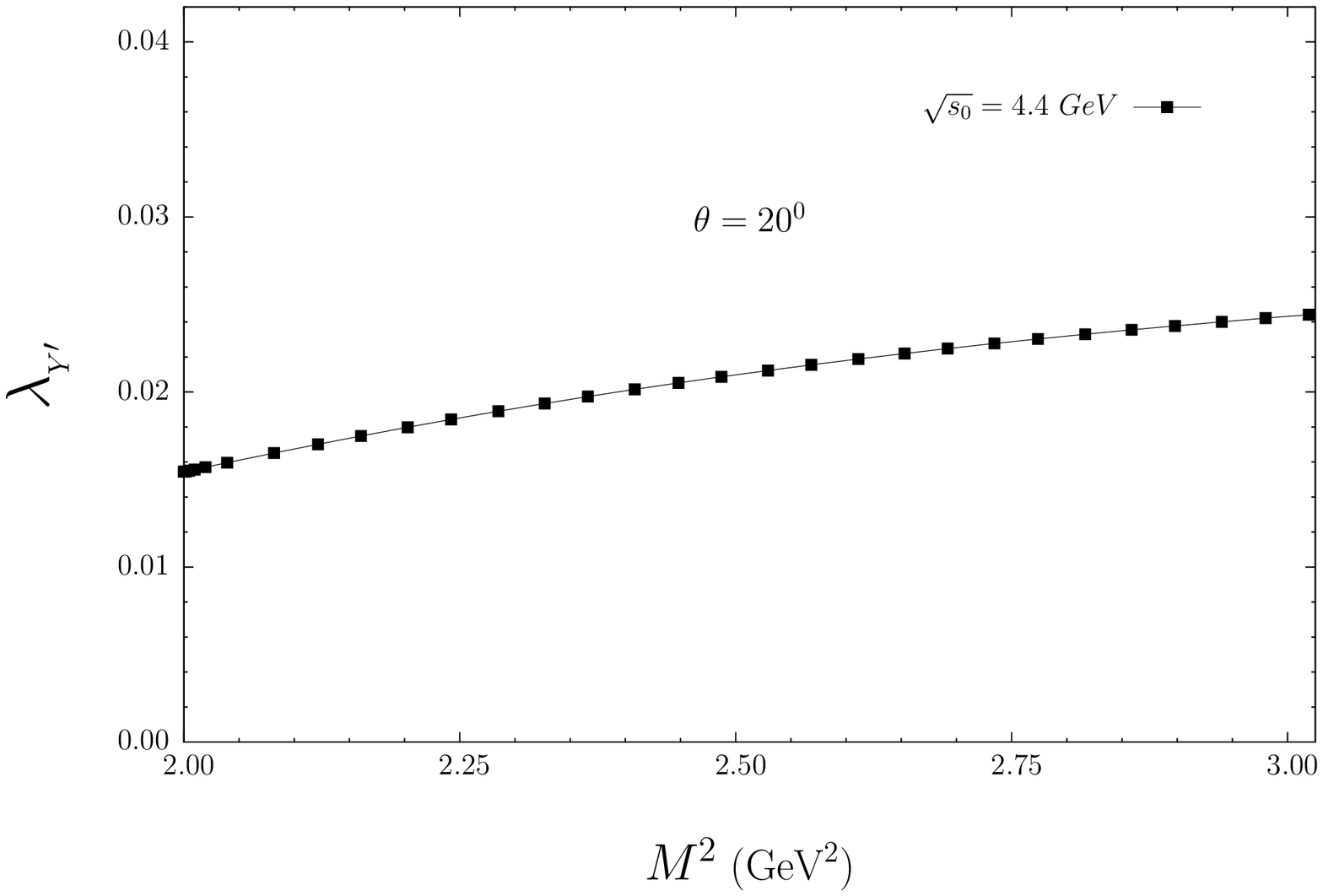}
\vskip 7.0cm
\caption{}
\end{figure}

\begin{figure}
\vskip 3. cm
    \includegraphics{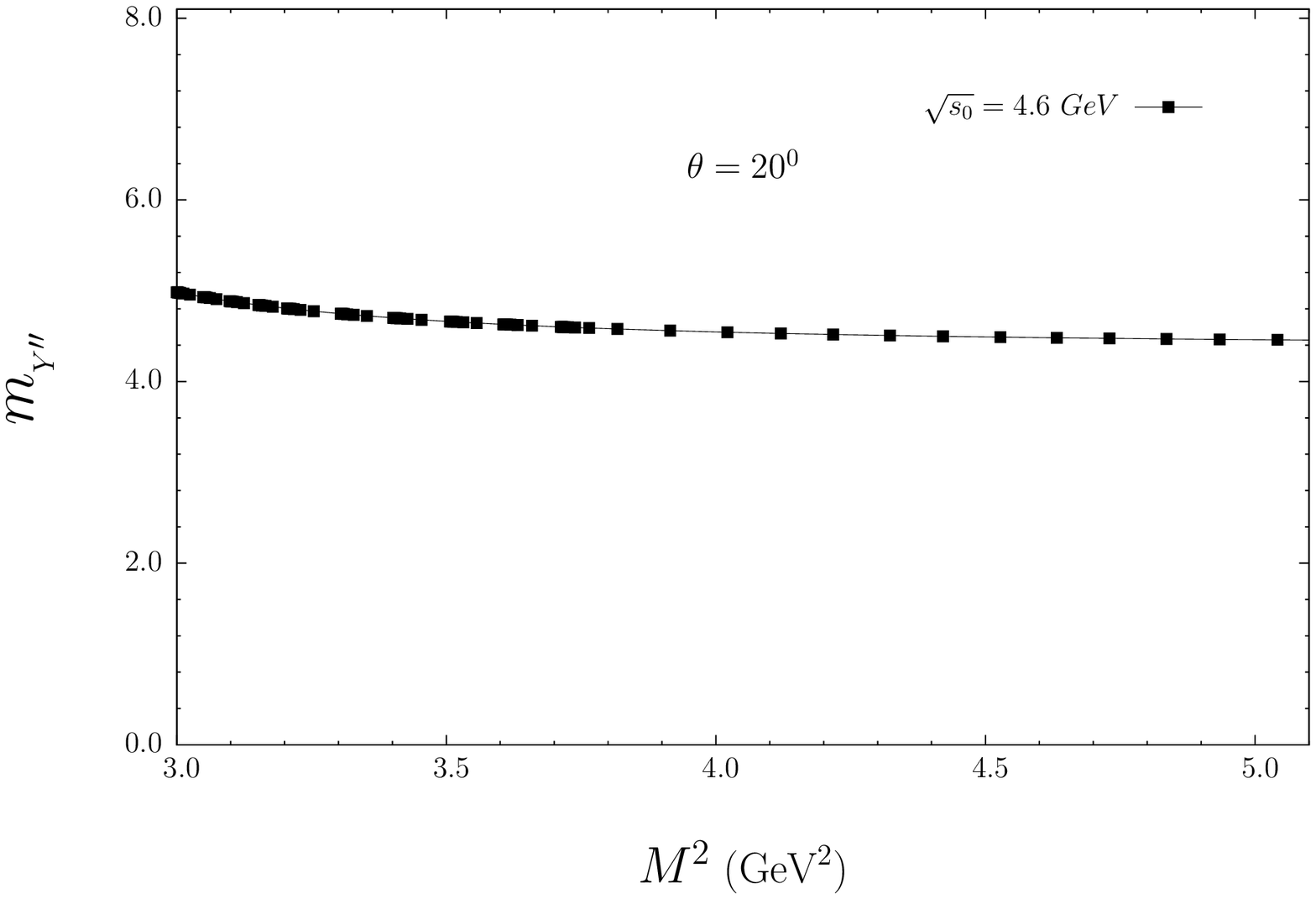}
\vskip 7.0cm
\caption{}
\end{figure}

\begin{figure}
\vskip 3. cm
    \includegraphics{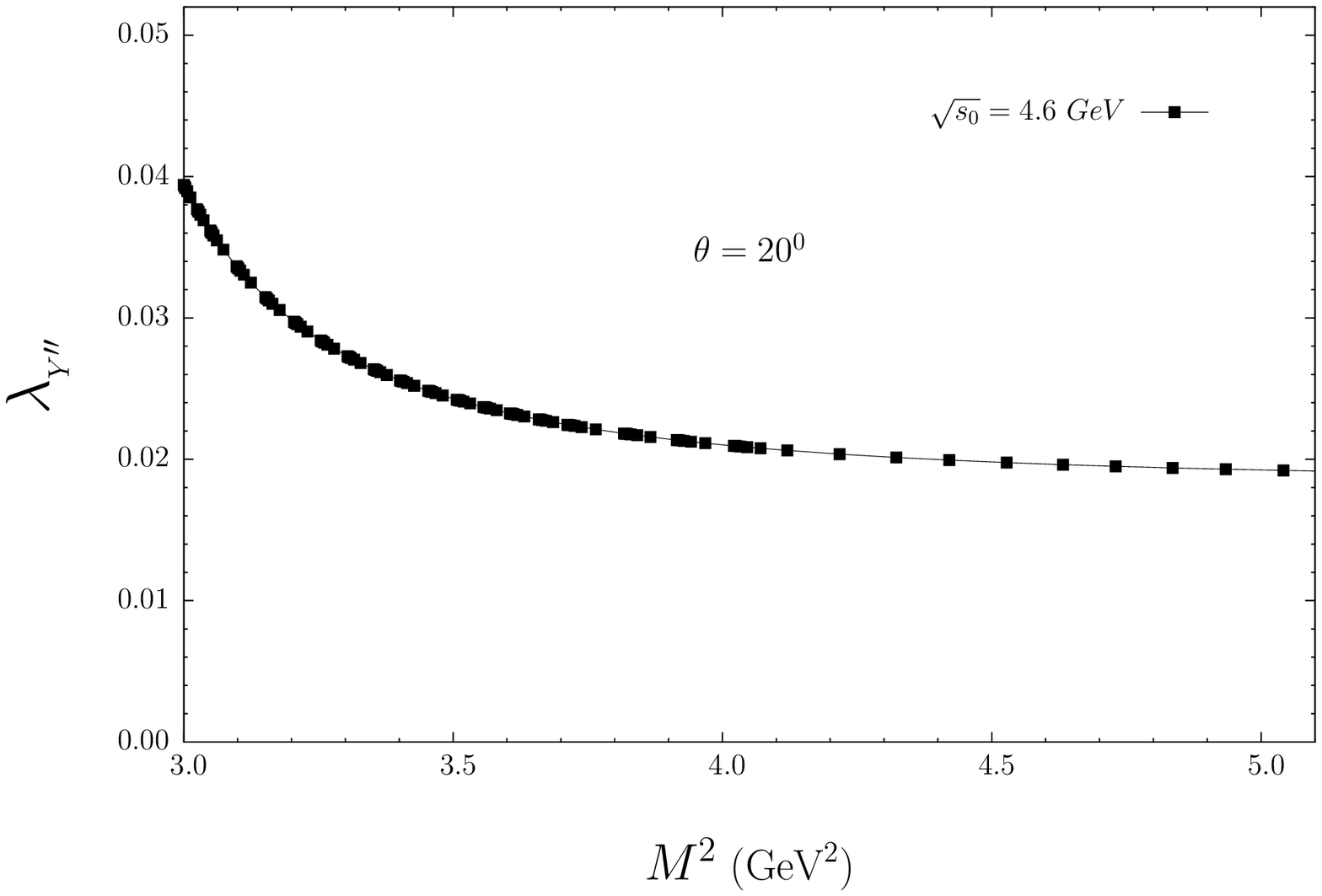}
\vskip 7.0cm
\caption{}
\end{figure}

\end{document}